\documentclass[twocolumn,showpacs,superscriptaddress,showkeys,preprintnumbers,amsmath,amssymb,prl]{revtex4}
\usepackage{graphicx}
\usepackage{dcolumn}
\usepackage{bm}
\usepackage{epstopdf}
\usepackage{color}

\begin{document}

\title{Criticality in vibrated frictional flows at finite strain rate}

\author{Geert Wortel}
\affiliation{Huygens-Kamerlingh Onnes Lab, Leiden University, PObox 9504, 2300 RA Leiden, The Netherlands}

\author{Olivier Dauchot}
\affiliation{EC2M, UMR Gulliver 7083 CNRS, ESPCI ParisTech, PSL Research University, 10 rue Vauquelin, 75005 Paris, France}

\author{Martin van Hecke}
\affiliation{Huygens-Kamerlingh Onnes Lab, Leiden University, PObox 9504, 2300 RA Leiden, The Netherlands}
\affiliation{FOM Institute AMOLF, Science Park 104, 1098 XG Amsterdam, The Netherlands}

\date{\today}

\begin{abstract}
We  evidence critical fluctuations in the strain-rate of granular flows that are weakly vibrated. Strikingly, the critical point arises at {\em finite} values of the mean strain rate and vibration strength, far away from
the yielding critical point at zero flow rate. We show that the global rheology, as well as the amplitude and correlation time of the fluctuations, are consistent with a mean-field, Landau like description, where strain rate and stress act as conjugated variables. We introduce
a general model which captures the observed phenomenology, and
argue that this type of critical behavior generically arises when self fluidization competes with friction.
\end{abstract}

\maketitle

Fluctuations play an essential role in flows of disordered media~\cite{Hebraud:1998vo,GDR,behringer,gutfraind,
lydericnature,foamcouette,kiri,reddy,CR,dijksman,wortelpre2,schall,clement,ruiz,pouliquen,kamrin,Bruno}.
In the simplest scenario, such fluctuations are rate-independent, as in thermal systems or strongly vibrated granular flows~\cite{clement,ruiz,jia}.
New phenomena, such as nonlocal rheology, arise when fluctuations are generated by the flow itself, as observed for  emulsions~\cite{lydericnature}, foams~\cite{foamcouette}, and granular matter~\cite{kiri,reddy,CR,dijksman,kamrin,pouliquen,Bruno}.
Granular media are particularly susceptible to flow-generated and externally provided fluctuations. First, the particles are so hard that tiny motions cause large fluctuations in the contact forces ~\cite{umbanhowar,CR}. Second, sliding friction is nearly rate independent, allowing subtle self-fluidization effects to qualitatively modify the slope of the flow curve, e.g. from neutral to negative~\cite{heinrich,hatano,dijksman}. Indeed, self-fluidization is particularly spectacular for granular media: for example, the finite yield threshold, a hallmark of static granular media, completely vanishes in the presence of flow {\em anywhere} in the granulate~\cite{kiri,pouliquen,kamrin,Bruno,dijksman}.

\begin{figure}[t!]
\begin{center}
\includegraphics[width=\columnwidth]{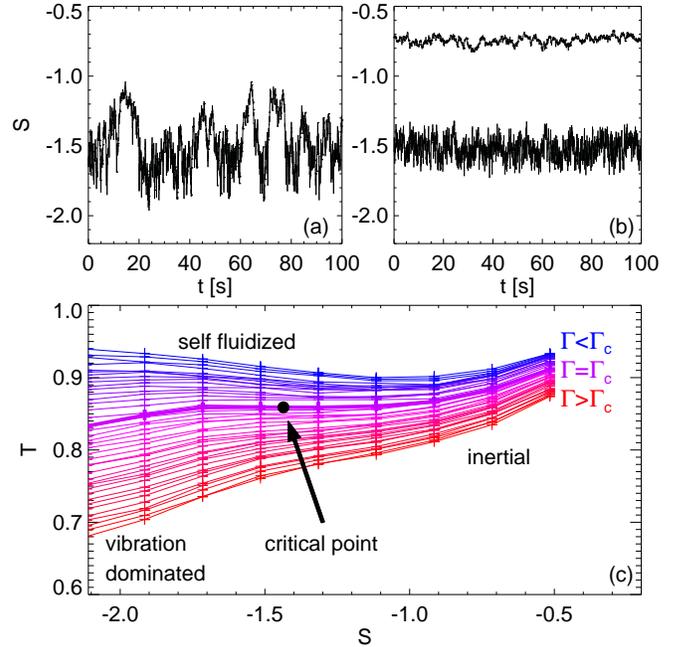}
\caption{(color online) Fluctuations and mean flow, characterized by $S:= log(I)$, where $I$ is the inertial number~\cite{noteI}.
(a-b) Flow rate fluctuations $S(t)$ detected in stress controlled experiments. (a) Strong and slow fluctuations close to the critical point ($(T,\Gamma)=(0.84,0.71)$). (b) Small fluctuations away from the critical point (top: $(T,\Gamma)=(0.85,0.71)$, bottom: $(T,\Gamma)=(0.79,0.92)$). (c) Flow curves measuring stress $T$ as function of flow rate $S$
for a range of  $\Gamma$ around $\Gamma_c$. The marginal flow curve at $\Gamma=\Gamma_c$ and corresponding critical point $(T_c,S_c)$ are indicated.}
\vspace{-12mm}
\label{rawflowcurves}
\end{center}
\end{figure}

What is the precise role of fluctuations for granular flows? Can local fluctuations organize in strong and slow collective fluctuations? How can we model the mutual coupling between fluctuations and flow?

To answer these questions, we probe the fluctuations and flow of weakly vibrated granular media sheared in a split-bottom cell~\cite{kiri,dijksman,splibo}.
First, by controlling the driving torque $T$
at finite shaking strength $\Gamma$, and
 measuring the time-resolved global flow rate $S(t)$, we reveal that fluctuations in $S$ become increasingly large and
slow when $(T,\Gamma) \rightarrow (T_c,\Gamma_c^+)$ (Fig.~1a-b). Second, we show that rheological curves  $T(S)$, obtained at fixed $S$,
and their  variation  as a function of  $\Gamma$ (Fig.~1c), can be captured in a mean-field type expansion around $(T_c,\Gamma_c)$. Together,
these experiments evidence the existence of a finite flow-rate (FFR) critical point.
While strong fluctuations have been studied near the zero flow-rate yielding point~\cite{behringer}, we stress that critical fluctuations at  FFR critical points have not been reported before, presumably because they remain hidden in absence of an external source of vibrations. Finally, we introduce a general model that combines a microscopic frictional rheology
with fluctuations of the microscopic stresses. This model successfully describes the experimentally observed $\Gamma$-dependent rheology and the emergence of the FFR critical point, naturally capturing the intricate coupling between stress, flow rate, and fluctuations.
Our results suggest that the FFR critical point is robust, and that
similar critical behavior may arise in  other frictional or nearly rate-independent systems, leading to potentially hazardous fluctuations  in previously overlooked flow regimes.

{\em Setup and phenomenology:} The vertically vibrated split-bottom cell has been described in detail previously \cite{splibo,dijksman,wortelpre2,splibonote}. In this system we drive granular flow by rotation of a disk and probe the driving torque $\tau$ and rotation rate $\Omega$ by a rheometer (Anton Paar DSR 301), which can be employed in rate or stress controlled modes.
We vertically vibrate the system as $A \sin(2\pi f t)$, with $f=63$~Hz, and control the dimensionless vibration strength $\Gamma=A(2\pi f)^2/g$, where $g$ is the gravitational acceleration.
The rheometer and vibrating flow cell are coupled through a flexure, and to accurately probe the disk rotation we use an optical angular encoder (Heidenhain ERO 2500) directly coupled to the disk.
We express our results in dimensionless units $T:=\tau/\tau_y$, where $\tau_y$ is the dynamic yield torque in the absence of external vibrations, and $S:= \log(I)$, where $I$ is the inertial number defined for pressures and strain rates at half depth \cite{noteI}.

{\em Critical Fluctuations:} We first perform experiments at constant torque $T$ and vibration amplitude $\Gamma>\Gamma_c$ and determine the magnitude and correlation time of the fluctuations in flow rate via the instantaneous angular position $\theta(t)$ of the bottom disk. We extract the rotation rate $\omega(t):=\partial_t \theta(t)$, after carefully checking that $\theta(t)$ is probed at sufficiently high temporal resolution. We then compute the averaged flow rate $\Omega = \left< \omega\right>$, the amplitude of its fluctuations $\sigma_{\omega}^2 = \left < \delta\omega^2 \right > $ and the temporal correlations $R(\tau) = \left < \delta\omega(t+\tau) \delta\omega(t)\right > / \sigma_{\omega}^2$, where $\delta\omega = \omega - \Omega$ and $\left < \cdot \right >$ denote temporal averages.  The correlation time $\tau_c$ is extracted by fitting the autocorrelation to an exponential (and is consistent with the time obtained by integrating the correlation function).

\begin{figure}[t]
\begin{center}
\includegraphics[width=\columnwidth]{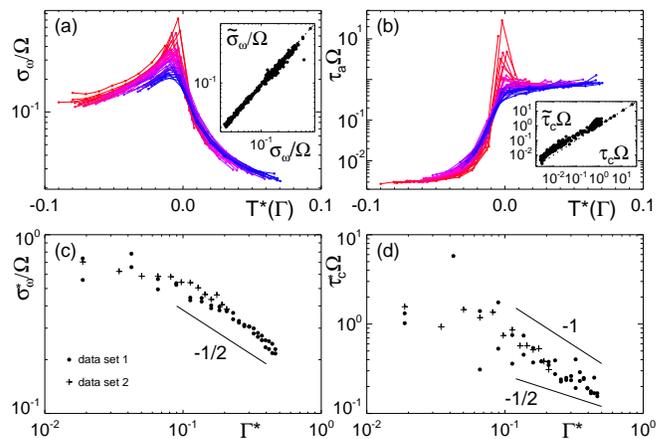}
\caption{(color online) Critical fluctuations. (a-b) Fluctuation magnitude $\sigma_{\omega}/\Omega$ as function of $T^*$ for $\Gamma$ from 0.65 to 0.94 --- red (peaked) curves have
 $\Gamma$ close to $\Gamma_c$. Inset: $\tilde\sigma_{\omega}$ and $\sigma_{\omega}$ are essentially equal. (b) Correlation time $\tau_c\Omega$ as function of $T^*$. Inset $\tilde\tau_c$ scales linearly with $\tau_c$. (c)-(d) Evidence for critical scaling of respectively $\sigma_{\omega}^{*}$ and $\tau_c^{*}$ with $\Gamma^*$ in two datasets taken several weeks apart.}
\label{flucts}
\end{center}
\vspace{-5mm}
\end{figure}

Fig.~\ref{flucts}a-b display the resulting dimensionless fluctuation amplitude $\sigma_{\omega}/\Omega$ and dimensionless correlation time $\tau_c \Omega $ as a function of the relative torque $T^*(\Gamma)=\left(T-T_i(\Gamma)\right)/T_i(\Gamma)$, where $T_i(\Gamma)$ is the inflection point in the flow curves, for different values of $\Gamma>\Gamma_c$. There is a sharp contrast between the fluctuations at either side of the peaks, which we interpret as signaling two qualitatively different flow regimes: a vibration dominated creep regime (strong but short-time correlated fluctuations) and a fast inertial flow regime (small fluctuations with time scale $\approx \Omega^{-1}$). Crucially, there is
a sharp transition between these regimes: both the fluctuation amplitude $\sigma_{\omega}/\Omega$ and correlation time $\tau_c \Omega $ exhibit a sharp maximum at $T^*\simeq 0$, which rapidly grows
when $\Gamma^*=(\Gamma-\Gamma_c)/\Gamma_c \rightarrow 0^+$.

To check the robustness of our measurements, we have also determined the fluctuation magnitude and correlation by considering the rotating disk as a massive random walker with drift.
We thus characterize the mean square angular displacement  $\Delta\theta(\tau)^2=\left <(\theta(t+\tau) - \theta (t))^2\right>$. The amplitude of the flow rate fluctuations $\tilde\sigma_{\omega}$ and the correlation time $\tilde\tau_c$ are then extracted from the asymptotics: for $\tau/\tilde\tau_c << 1$, we observe ballistic dynamics with $\Delta\theta(\tau)^2 \sim \tilde\sigma_{\omega}^2 \tau$, while for $\tau/\tilde\tau_c >>1 $ the dynamics is diffusive, with $\Delta\theta(\tau)^2 \sim 2 \tilde\sigma_{\omega}^2 \tau_c \tau$. These two independent protocols yield consistent results, as shown in the insets of Fig.~\ref{flucts}ab.

The peak in the flow rate fluctuations diverges in a manner consistent with a power-law scaling $\sim {\Gamma^*}^{-\tilde\gamma}$, with $\tilde\gamma \approx 0.5$  (Fig.~\ref{flucts}c).
The correlation times are too noisy to be reliably fitted to a power law ${\Gamma^*}^{-\mu}$, but if any, $\mu\in[0.5, 1]$ (Fig.~\ref{flucts}d).
Together, these signals provide strong evidence for  critical behavior.

\begin{figure}[t]
\begin{center}
\includegraphics[width=\columnwidth]{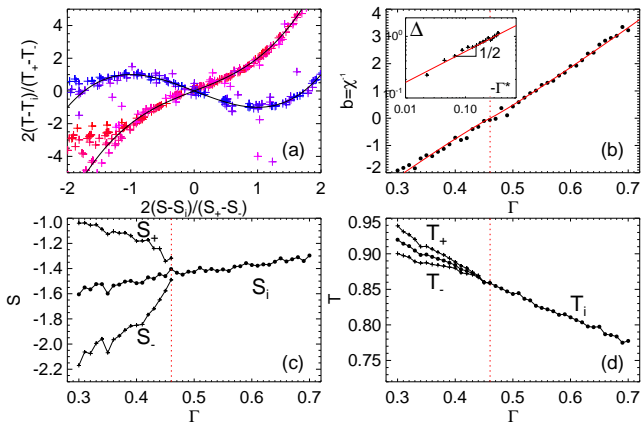}
\caption{(color online) (a) Flow curve data collapsed onto master curves. (b) Inverse susceptibility $\chi^{-1}$ vs. $\Gamma$. Inset : log-log plot of $\Delta = S_+ - S_-$  as a function of $-\Gamma^*$. (c-d) Location of the extrema $(S_{\pm},T_{\pm})$ and the inflection point $(S_i,T_i)$ in the planes $(\Gamma, S)$ and $(\Gamma, T)$. The vertical dashed line indicates $\Gamma_c$.}
\label{flowcurves-fit}
\end{center}
\vspace{-7mm}
\end{figure}

\emph{Scaling of the flow curves:} The critical behavior reported above suggest that the torque and flow rate should be related via a Landau type expansion in the critical regime:
\begin{equation}
T= a (S-S_i)^3 + b (S-S_i)+T_i~, \label{fit}
\end{equation}
where $(T_i(\Gamma),S_i(\Gamma))$ are the inflection points of the flow curves. In order to probe this relation, we perform rate controlled experiments, in which we can also access the negative slope regime. The flow curves, shown in Fig.~\ref{rawflowcurves}c, are indeed reminiscent of a third order polynomial. Fitting the data accordingly, we extract $(S_i,T_i)$, $a$ and $b$, and the local maximum $(S'_{+},T'_{+})$ and minimum $(S'_{-},T'_{-})$ as a function of $\Gamma$.
As shown in Fig.~\ref{flowcurves-fit}a, the flow curves can be rescaled on two distinct branches, below and above $\Gamma_c$, over a substantial range. As expected, the cubic coefficient $a$ remains essentially constant ($a \simeq 2$, not shown here). The coefficient $b$, which sets the slope at the inflection point, akin to an inverse susceptibility $\chi^{-1}$, crosses zero at $\Gamma=\Gamma_c$ and increases linearly with $\Gamma^*$.
The location of the extrema $(S_{\pm},T_{\pm})$ in the $(S,\Gamma)$ and $(T,\Gamma)$ planes, displayed on Figs~\ref{flowcurves-fit}c-d, together with the location of the inflection point $(S_i,T_i)$, determine the so-called spinodal lines, which are the stability limits of the fast and slow flow phases. The region of  "coexistence" corresponds here to the set of parameters for which the flow curves have a negative slope. The width of this region $\Delta = S_+ - S_-$ scales like $|\Gamma^*|^{\beta}$, with $\beta = 0.5$ for $\Gamma^*<0$, in agreement with Eq.~(\ref{fit})  and the linear dependence of $b$ with $\Gamma^*$.

{\em FFR Critical Point:}  Our data for both stress-controlled and strain-rate controlled experiments provide strong evidence for the existence of a critical point at finite flow rate, characterized by the following scaling relations:
\begin{eqnarray}
\Delta \sim& {\Gamma^*}^{\beta}; \quad \beta &\simeq  0.5 \\
\chi \sim& {\Gamma^*}^{-\gamma}; \quad \gamma &\simeq  1 \\
\sigma_\omega/\Omega \sim& {\Gamma^*}^{-\tilde\gamma}; \quad \tilde\gamma &\simeq  0.5 \\
\tau_c \Omega \sim& {\Gamma^*}^{-\mu}; \quad \mu & \in [0.5, 1] .
\end{eqnarray}
We note that in stress control experiments, $\Gamma_c=0.65\pm 0.01$, as determined from both the zero slope inflection point of the flow curves, and the diverging fluctuations, while in strain rate controlled experiments, $\Gamma_c = 0.46 \pm 0.01$, as determined from the zero slope inflection point of the flow curves. We believe this difference to be due to the complex combination of large intrinsic fluctuations, finite size effects, and the non-perfect feedback loop of the rheometer in rate controlled experiments.

The fact that the critical behavior of $\chi$, obtained for averaged quantities in strain rate controlled experiments, and that of $\sigma_{\omega}^2$, obtained from the fluctuations in stress controlled experiments, coincide is a strong indication of the relevance of our analysis. Both the value of the exponents and the quality of the description of the flow curves by Eq.~(\ref{fit}) suggest that a mean field description should capture the essence of the observed phenomenology.

{\em Flow Model:}
We finally introduce a general fluctuation-frictional (FF) model that captures the observed rheology. We combine a frictional local rheology with fluctuations that are induced by both vibrations and flow, and show that the average rheology of this model exhibits all the experimentally observed hallmarks, including the FFR critical point. First, we introduce an \emph{agitation strength} $A$, which is a function of both
flow-induced and vibration-induced agitations \cite{footnoteIS}:
\begin{equation}
A=A_g(\Gamma,I)~, \label{agi}
\end{equation}
where $A_g=0$ only when both $\Gamma$ and $I$ are zero. Second,
 we postulate that the local stresses $T_m$ are fluctuating around their mean $T$, and that the
microscopic stress distribution $P(T_m)=\bar{P}((T_m\!-\!T)/A)$, where $\bar{P}(x)$ is a given normalized distribution centered at $x=0$ --- note that $A$ sets the width of $P(T_m)$. Third,  we determine
the global flow rate $I$ as the mean of the microscopic flow rates $I_m$, where
$I_m$ and $T_m$ are related by simple frictional Herschel-Bulkley rheology with a finite yield stress --- in particular, $I_m=0$ when $|T_m|<1$. Combining these ingredients, we find:
\begin{equation}
I(A,T)=\!\int_{-\infty}^{\infty} \frac{dT_m}{A} ~ ~ \bar{ P}\left(\frac{T_m-T}{A}\right) ~ I_m(T_m)  ~.
 \label{Ieqmain}
\end{equation}
For a prescribed set of functions $A_g, P$ and $ I_m(T_m)$, Eqs.~\ref{agi}-\ref{Ieqmain} completely set the flow curves $T(I,\Gamma)$.
We start with a definite choice of $A_g$, $\bar{P}$ and $I_m$ \cite{detailnote}
and then show that our conclusions are insensitive to this choice - details are provided in the Supp. Mat.~\cite{sup}.

The FF model exhibits
the experimentally observed  singularity of the stress $T(I,A)$ at the origin:
when $A=0$,
Eq.~(\ref{Ieqmain}) implies that
$T_m=T$, $I_m=I$, so that the macro rheology is identical to the microscopic rheology and $T(0+,0)=1$.
In contrast, when $A>0$ but $I=0$,
Eq.~(\ref{Ieqmain}) implies that the stress distribution must be symmetric around zero --- therefore $T=0$, and in particular
$T(0,0^+)=0$. The model thus captures  the discontinuous
vanishing of the yield stress when $\Gamma$ becomes finite.

\begin{figure}[t]
\begin{center}
\includegraphics[width=\columnwidth,clip,trim=-.4cm -.8cm .3cm 3.5cm]{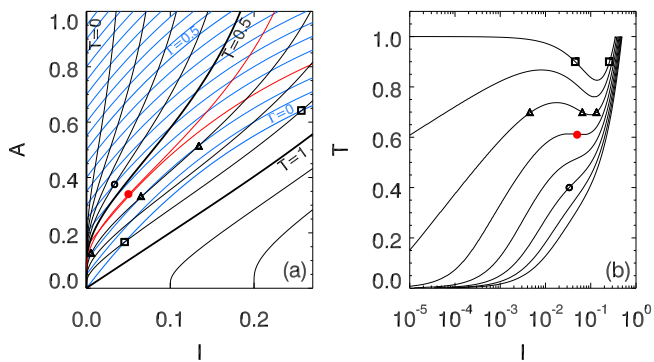}
\caption{(color online) FF Model - for details see \cite{detailnote}.
(a) Contour curves of constant $\Gamma$ (blue)
and constant $T$ (black).
Their intersections, for example
at $(\Gamma,T)=(0.25,0.4)$ (open circle),
$(0.1,0.7)$ (three triangles), and $(0,0.9)$ (two squares) determine $T(I)$.
The red contour curves for $T =0.6$ and $\Gamma = 0.15$
are (nearly) tangent, leading to the critical point
(red filled circle).
(b) Corresponding flow curves for $\Gamma=0,0.05,\dots, 0.35$ (ordered from top to down) ---
circles, squares and boxes match those in panel (a).}
\label{model}
\end{center}
\vspace{-7mm}
\end{figure}

We now show that the FF model captures all qualitative features of the rheology of weakly vibrated flows. The solutions to this model can be understood by considering
the variation of  $\Gamma$ and $T$ in the  $(I,A)$-plane (Fig.~\ref{model}).
The flow curves $T(I,\Gamma)$ can be determined graphically via
the intersections of the contour curves of $T$ and $\Gamma$, by
fixing $\Gamma$, varying $T$, and determining the corresponding value(s) of $I$.
For $T>1$ there is only one intersection, corresponding to rapid flows, and in the remainder we focus on $T\le 1$. {\em (i)} For large $\Gamma$, there is only one intersection (black dot), leading to monotonic flow curves $T(I)$. {\em (ii)} For small $\Gamma$, there are three intersections (crosses), corresponding to non monotonic flow curves. {\em (iii)}
In between these two regimes is the critical $\Gamma_c$ curve (red), for which the three intersection points merge (red dot). {\em (iv)} Finally, for $\Gamma=0$, there are precisely
two intersections, corresponding to the only flow curve that has
finite $T$ and negative slope at $I=0$.
We stress that the scenario that emerges captures the essence of the experimental flow curves shown in Fig.~1a, without
having to make any assumptions about the behavior of $A$ near the FFR critical point.

Clearly, the essence of this scenario does not depend on the details of the agitation function $A_g$, distribution $\bar{P}$ and local rheology $T_m$. The only condition is that
the $\Gamma=0$ curve is steeper than the $T=1$ contour at the origin, such that there are two intersections between the $\Gamma=0$ and $T=1$ curves --- for other examples of flow curves, including for when this condition is violated,  see the Supp. Mat.~\cite{sup}.

\emph{Discussion:} We have uncovered a dynamical critical point in agitated frictional flows.
We stress again that the concomitant large fluctuations arise at finite flow rates, away from the yielding point where strong fluctuations have been seen before~\cite{behringer}, which makes this an experimentally easily accessible yet nontrivial critical point which deserves further investigation.
We argued that this criticality emerges from the interplay of external vibration and self-fluidization. Both act as sources of agitation, but influence the rheology very differently: while external agitations set the yield stress to zero and impose a positive slope in the limit of zero flow rate,
flow induced fluctuations cause a negative slope in the flow curves, at least in the absence of externally provided fluctuations. The FFR critical behavior emerges due to the competition of these.

From a more theoretical point of view,
the model we have introduced here is purely phenomenological and essentially mean field. An alternative, more complete strategy would be to write down a dynamical equation for the microscopic stress distribution, as introduced in the H\'ebraud-Lequeux~\cite{Hebraud:1998vo}  and related fluidity models~\cite{Bocquet:2009kt,Mansard:2011cw}, which at present do not capture
the critical scaling of the flow curves nor the diverging fluctuations.
In these models the self-fluidization is captured by a diffusion term for the local stresses, the amplitude of which is linearly related to the amount of stress exceeding the local yield stress.
Adding any finite amount of external noise via a constant term in the diffusion amplitude, we expect
the dynamical yield stress to vanish as observed here. Then, following~\cite{Mansard:2011cw}, one could work out the relation between model parameters and flow rate, and hopefully obtain the observed non-monotonic rheology. Stress fluctuations could also be taken into account following the very recent work by Agoritsas et al.~\cite{Agoritsas:2015wj}. It is an open question whether such models can exhibit a negative slope at zero flow rate in the absence of external vibrations.

\noindent\emph{Acknowledgments --} We thank J. Dijksman, J. Mesman, H. Eerkens,
E. Agoritsas, E. Bertin and K. Martens for discussions, technical support, and early experiments, and FOM/NWO for funding.

\newpage
~\newpage

{\em \Large Supplemental Material }

\begin{figure}[t]
\begin{center}
\includegraphics[width=0.7\columnwidth,clip,trim=5cm -1.cm 0.5cm 3.5cm]{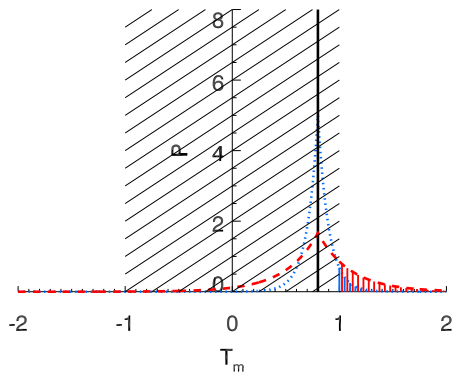}
\caption{Probability distribution for the stresses $T_m$, for $T=0.8$, $A\rightarrow 0$ (line),
$A=0.1$ (dotted, blue) and $A=0.3$ (dashed, red). The diagonally hatched box indicates the jammed region where $|T_m|<1$,
and the hatched tails of $P(T_m)$ outside this box indicate the weight of $P(T_m)$ that contributes to flow.
 }
 \label{idea}
\end{center}
\end{figure}

Here we provide a detailed discussion of the FF model that relates fluctuations, flow and stress.
The aim of this model is to explore how the main experimental findings for slow, i.e., non-inertial, weakly vibrated granular flows can be captured by introducing a scalar
agitation strength $A$. We note
that the experimentally observed increase of $T(I)$ for large values of $I$ is due to inertial effects
that our model does not necessarily capture.

The central idea of the FF model is illustrated in Fig.~\ref{idea}.
We assume that in the presence of agitations of magnitude $A$, a macroscopically applied stress $T$ leads to a distribution of microscopic  stresses $P(T_m)$. We assume $P(T_m)$ to be symmetric, peaked around $T_m=T$, and with a width proportional to $A$ ---  in the limit of vanishing agitation strength, $P(T_m)$ approaches a $\delta$-function (Fig.~\ref{idea}). For definiteness, we assume here that $P$ has exponential tails, although this is not essential for our qualitative picture:
\begin{equation}
P(T_m)=(2A)^{-1} \exp(-|(T_m-T)|/A)~.
\label{SIP}
\end{equation}

We furthermore assume that a local rheology relates
a distribution of flow rates $I_m$ to $P(T_m)$, and that the
global flow rate $I$ is the mean of $I_m$:
\begin{equation}
I(A,T)=\!\int_{-\infty}^{\infty} dT_m ~{ P}\left(T_m\right) ~ I_m(T_m)  ~.
\label{SIint}
\end{equation}
For the local rheology we take
a  Herschel-Bulkley form with unit yield stress, i.e.,
\begin{equation}
T_m=\mbox{sign}(I_m) \left[1+|I_m|^{\alpha}\right]~,
\label{SIHB}
\end{equation}
where for simplicity we set $\alpha=1$. This captures the essential
feature of a microscopic flow threshold, as $I_m=0$ in the jammed region $|T_m|<1$.

An important consequence of Eq.~(\ref{SIP}-\ref{SIHB}) is
the singular difference in global flow rheology between $A=0$ and $A\ne 0$. When $A=0$, the global flow follows the local rheology, and there only will be flow when $|T|>1$. However, when $A\ne0$,
$P(T_m)$ has a finite weight outside the jammed region, and $I$ will be finite for any value of $T$,
unequal to zero. In other words, $\lim_{A\rightarrow 0} (\lim_{I \rightarrow 0} T) = 0$
 $\lim_{I\rightarrow 0} (\lim_{A \rightarrow 0} T) = 1$.

\begin{figure}[t]
\begin{center}
\includegraphics[width=\columnwidth,clip,trim=0cm -.2cm 1.6cm 3.4cm]{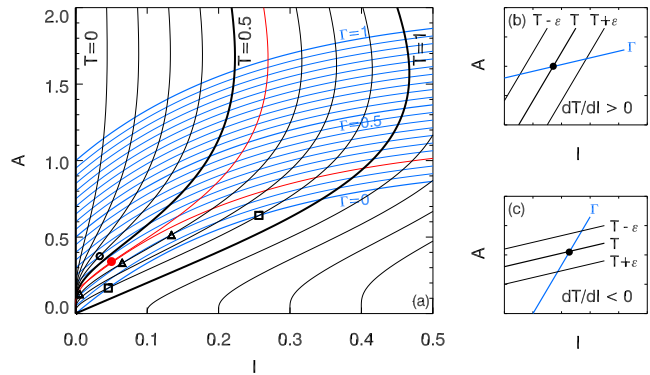}
\caption{(a) $T$-contours and $\Gamma$-contours in the $(A,I)$-plane determined in the FF model,
using Eq.~(\ref{expa}) for the agitation strength. Colors and points are identical to those in Fig.~4 of the main text. (b-c) Schematic representations showing the intersections between $T$-contours and a $\Gamma$-contour for two different slopes of the rheological curve $T(I)$.
 }
 \label{SIZO}
\end{center}
\end{figure}

What sets the agitation strength? Clearly, $A=0$ when both $\Gamma$ and $I$ are zero, and
$A$ should be non-zero
when either $I$ or $\Gamma$ are finite. This strongly suggests that $A_g(\Gamma,I)$ is linear
in both $\Gamma$ and $I$ for small values of these arguments. Below, we show the results for two functional forms of $A$. The simplest linear agitation function reads:
\begin{equation}
A_g(\Gamma,I)= \kappa \Gamma + \lambda I~,
\label{lina}
\end{equation}
where we can set $\kappa=1$ by overall scaling of $A$, and where we will show $\lambda$ to be an important parameter.
In the main text, we have taken a slightly more complex form for  $A_g(I, \Gamma)$:
\begin{equation}
A_g(\Gamma,I)= \Gamma+(1-\exp(I/I_0)), \mbox{ with } I_0=0.25~,
\label{expa}
\end{equation}
where the exponential form is motivated by the observation that
the flow-induced fluctuations likely saturate at large flow rates.

{\em Qualitative Properties:} Even without solving the coupled equations (\ref{SIP}-\ref{expa}), most qualitative properties
of their solutions can be shown in a straightforward manner to match our experimental findings.
The first two properties follow from the singular difference in global flow rheology between $A=0$ and $A\ne 0$ discussed above, and capture the singular vanishing of the yield stress when $\Gamma$ becomes finite:\\
{\em (1)} When $\Gamma \ne 0$ and $T \ne 0$, there is flow. \\
{\em (2)} When $\Gamma=0$, the system is jammed for $|T|<1$.\\
A corollary of the first property is that when $\Gamma \ne 0$, the stress smoothly goes to zero when $I\rightarrow 0$, implying that the flowcurve $T(I)$ has a positive slope for small $I$.

\begin{figure}[t]
\begin{center}
\includegraphics[width=\columnwidth,clip,trim=-.3cm -.7cm .4cm 3.3cm]{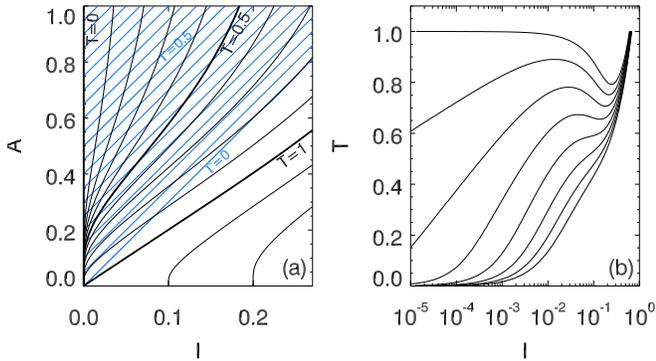}
\caption{Contourplots (a) and rheological curves (b) for the FF model, for a linear agitation function
$A_g(\Gamma,I)= \Gamma+3\times I$
 }\label{l3}
\end{center}
\end{figure}

The slope of the flow curve for $\Gamma=0$ requires a more subtle analysis, involving the rheological curves. These rheological curves $T(I)$ for fixed $\Gamma$ can be obtained graphically, by considering both $T$ and $\Gamma$ as a function of $I$ and $A$. To do so, we fix $T$ and vary $A$ to obtain $I$ using Eq.~(\ref{SIint}), which yields contours of fixed $T$,
and use Eq.~(\ref{lina}) or (\ref{expa}) to plot $A(\Gamma,I)$ at fixed $\Gamma$ yielding $\Gamma$-contours.
In Fig.~\ref{SIZO}a we show these curves, for the agitation function given by Eq.~(\ref{expa}).
Fixing $\Gamma$, the rheology, i.e. $I(T)$ follows from intersections of the pertaining $\Gamma$-contours and $T$-contours, where we note that there may be multiple solutions.
Fig.~\ref{SIZO}b-c schematically illustrate that it is the relative slope of the $\Gamma$-contours and $T$-contours which determines whether the flow curve $T(I)$ has a positive or negative slope.

\begin{figure}[t!]
\begin{center}
\includegraphics[width=\columnwidth,clip,trim=-.3cm -.7cm .4cm 3.3cm]{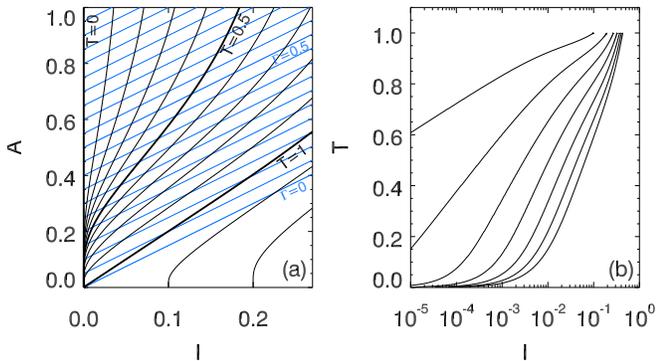}
\caption{Contourplots (a) and rheological curves (b) for the FF model, for a linear agitation function
$A_g(\Gamma,I)= \Gamma+1.5\times I$.
 }\label{l15}
 \end{center}

 \end{figure}

In particular, the question whether the flow curve for $\Gamma=0$ has a negative slope for small $I$,
can be answered by inspecting the slopes of the
$\Gamma=0$ curve and the $T=1$ curve which meet at the origin (Fig.~\ref{SIZO}a). Expanding
Eq.~\ref{SIint} for small $A$ yields that $A \approx 2 I$ along the $T=1$ contour. Hence, as long as
$\partial_A I|I=0>2$ (evaluated along the $\Gamma=0$ contour), the $\Gamma=0$ flow curve will have a negative slope. This is manifestly true for the agitation function given by Eq.~(\ref{expa}), and
is also true for the agitation function given by Eq.~(\ref{lina}), provided that $\lambda >2$.
In Figs.~\ref{l3} and \ref{l15} we show examples of the contourplots and corresponding rheological curves for $\lambda=3$ and $\lambda=1.5$. The former clearly exhibits a $\Gamma=0$ flow curve with negative slope, and corresponding FFR critical point:  details of the agitation function are not important for the overall scenario. However, when $\lambda$ is too low, the
scenario changes qualitatively: even though the origin is still singular,
there is no longer a finite yield stress for $\Gamma=0$, no negative slope, and no FFR critical point (Figs.~\ref{l15}).


\begin{thebibliography}{99}
\bibitem{splibonote} Split-bottom flow geometries produce smooth, robust, and well controlled granular flows.
Our cell consists of an acrylic cylindrical container (inner radius 7.0~cm), at the bottom of which a rotating disk of radius $r_s$ of 4~cm
drives the flow. To ensure no-slip boundary conditions, the top surface of the disk and cylinder bottom is made rough. The cell is filled to a height $H$ of 24 mm with black soda-lime glass beads with diameters ranging from 1 to 1.3~mm.
All experiments are carried out under ambient temperature, pressure and relative humidity, and our experiments reproduce well over the course of several years --- more details in \cite{dijksman,wortelpre2}.
\bibitem{noteI} We use the standard definition of $I:=\dot{\gamma}d/\sqrt{P/\rho}$ \cite{GDR}, which, 
based on our estimates of typical values of $\dot{\gamma}$ throughout the flowing zone amounts to $I=(0.3 s) \Omega$ \cite{splibo}.
\bibitem{footnoteIS} We use $I$ rather than $S$ in the model, as we want to consider the absence of flow at $I=0$.
\bibitem{detailnote} We take $A_g(I, \Gamma)=\Gamma+(1-\exp(4 I))$ which
captures that both vibrations and flow induce fluctuations which saturate for large flow rates,
 $\bar{P}(x)=1/2\exp(-|x|)$, and $T_m$ follows
Herschel-Bulkley rheology:
$T_m=\mbox{sign}(I_m) \left[1+|I_m|\right]$, with $I_m=0$ when $|T_m|<1$.
\end{thebibliography}
\end{document}